
\def\gsim{\mathrel{\scriptstyle{\buildrel > \over \sim}}}
\def\lsim{\mathrel{\scriptstyle{\buildrel < \over \sim}}}
\magnification 1200
\overfullrule=0pt
\baselineskip=17pt

\centerline{\bf LAYERED XY-MODELS, ANYON SUPERCONDUCTORS,
AND SPIN-LIQUIDS}
\vskip 50pt
\centerline{J. P. Rodriguez}
\medskip
\centerline{{\it Dept. of Physics and Astronomy,
California State University,
Los Angeles, CA 90032}\footnote*{Permanent address.}}
\centerline{\it and Theoretical Division,
Los Alamos National Laboratory,
Los Alamos, NM 87545.}

\vskip 30pt
\centerline  {\bf  Abstract}
\vskip 8pt\noindent
The partition function of the double-layer $XY$ model
in the (dual) Villain form is
computed exactly in the limit of
weak coupling between layers.  Both layers are found
to be locked together through the Berezinskii-Kosterlitz-Thouless
transition,
while they become decoupled well inside  the normal phase.
These
results are recovered
in the general case of a finite number of
such layers.
When re-interpreted in terms
of the dual problems of lattice anyon superconductivity
and of spin-liquids, they also indicate that the
essential nature
of the transition into the normal state
found in two dimensions
persists in the case of a finite number
of weakly coupled layers.

\bigskip
\noindent
PACS Indices:  75.10.Hk, 74.20.Kk, 74.20.Mn, 11.15.Ha
\vfill\eject

The layered crystal structure common to high-temperature
superconductors is known to be reflected in their electronic
properties.$^1$  There are suggestions in
the theoretical literature that the effect of strong repulsive
interactions  present within a single such copper-oxygen
layer result in exotic anyonic superconducting states
and/or spin-liquid states.$^{2-5}$  Recently, the author
has shown that both of these groundstates undergo a phase
transition into a normal phase at non-zero temperature,
where the corresponding particle-hole excitations
become deconfined upon heating through the
transition.$^6$  The transition itself is dual to that
of    the two-dimensional (2D) Coulomb gas.$^{7,8}$  In the
present context of oxide superconductivity, therefore,
one is naturally led to the question of
how a finite number of such weakly coupled
layers behave.

In response to this query,
we shall study the problem of
a finite number of weakly coupled layers of square-lattice
$XY$-models in the Villain form,$^9$
which is known to lead to a dual theory
of (layered) compact quantum electrodynamics (QED)
in the strong-coupling limit.$^{10}$
The Abelian gauge-field that appears
here  describes chiral spin-fluctuations
in the case of the 2D spin-liquid state,
while it represents the statistical
flux-tube attached to each psuedo-fermion in the case of anyon
superconductivity.$^{3-5,11}$
Korshunov has given an approximate
solution of the $XY$-model with infinitely many
weakly coupled layers,
where he finds a
Berezinskii-Kosterlitz-Thouless (BKT)
transition within each layer,
followed by an inter-layer
decoupling transition at much higher
temperature that is characterized by
the binding of vortex rings (fluxons)
lying in between consecutive layers.$^{12}$
Note, therefore, that the lower-temperature
superfluid BKT transition
involves the {\it entire} three-dimensional (3D) lattice in this
scenario, which is consistent with previous
Monte Carlo simulations.$^{13}$
In this Letter, we first solve the
weakly coupled double-layer
$XY$ model in the Villain form {\it exactly}, recovering
Korshunov's approximate  results for the
case of infinitely many layers
in the processes.
The intra-layer phase auto-correlation function
of this double-layer $XY$-model is computed,
where we find that it falls off algebraicly
with distance below the BKT transition, as expected.
Notably, the exponent of the  correlation
function in question
vanishes linearly with temperature,
yet the coefficient is only half
as  big as that of
the single-layer case.$^9$
This indicates that the existence of neighboring
layers enhances long-range phase correlations, as
expected.
In addition, the inter-layer phase autocorrelation function
is computed, where we find that it coincides
with the prior intra-layer
phase autocorrelation function
at temperatures below the decoupling transition.
This, on the other hand, gives strong support to the
claim that the double-layers become locked
together at temperatures
below the latter transition.
Last, we show that the above double-layer results
persist in the general case of a finite number of
weakly coupled layers, suggesting that the superfluid transition
exhibited by the $XY$ model  with a finite number of
layers is 3D-like.$^{12,13}$

When the above  results are re-interpreted in terms
of the dual problem of double-layer anyon superconductivity
and spin-liquids, they yield that a confining
string exists between particle-hole excitations
at temperatures below the
deconfinement transition.$^{10}$  Note that the
consecutive layers have already become decoupled
at this stage via the
inter-layer
fluxon-antifluxon unbinding transition that occurs at
much lower temperature in this case.
Taking into account arguments given previously
by the author,$^6$ this implies that
the  basic   character of the single-layer particle-hole
deconfinement transition
into the normal state persists in the case
of double-layer anyon
superconductors and spin-liquids; i.e., the transition remains
2D.

Let us first consider the generalized correlation
function
$\langle {\rm exp} [i\sum_{r} p(r)\phi(r)]\rangle
= Z[p]/Z[0]$ of the $N$-layered $XY$-model, where the partition function
$Z[p]=\int {\cal D} \phi (r) e^{-E/T}$
has an energy functional
$$\eqalignno{{E\over T} = \beta_{\parallel}
\sum_{l=1}^{N}\sum_{\vec r}\sum_{\mu=x,y}
\{1-{\rm cos}[\Delta_{\mu}\phi(\vec r,l)]\}
&+\beta_{\perp}\sum_{l=1}^{N-1}\sum_{\vec r}
\{1-{\rm cos}[\phi(\vec r, l)-\phi(\vec r, l+1)]\} &\cr
&-i\sum_{l=1}^{N}\sum_{\vec r}p(\vec r, l)\phi(\vec r, l).
& (1) \cr}$$
Here $\vec r$ ranges over the square-lattice and $l$ denotes
the layer index, while $\beta_{\parallel,
\perp}=J_{\parallel,\perp}/k_B T$ is proportional
to nearest-neighbor
coupling-constants satisfying $J_{\parallel}\gg J_{\perp}$.
Also, $\Delta_{\mu}\phi(r)=\phi(r+\hat\mu)-\phi(r)$
is the lattice difference operator.
The particular correlation to be probed is determined
by the fixed integer field $p(\vec r, l)$; e.g., the choice
$p(\pm r/2, 0, 1)=\pm 1$ corresponds to the intra-plane
auto-correlation function.  After making the usual
(low-temperature) Villain substitution  for the exponential above,$^9$
$e^{-\beta [1 - {\rm cos}\,\theta]}\rightarrow
(2\pi\beta)^{-1/2}\sum_{n=-\infty}^{\infty}
e^{i n\theta} e^{-n^2/2\beta}$,   and then
integrating over the phase-field, we obtain the
following dual representation equivalent to
$N$-layered compact QED in the strong-coupling
limit (modulo a constant):$^{10}$
$$Z[p] = \sum_{\{n_{\mu}(r)\}}\Pi_{r}
\delta\Bigl[\Delta_{\nu} n_{\nu}|_{r}
-\sum_{r^{\prime}} p(r^{\prime})
\delta_{r^{\prime}r}\Bigr]
{\rm exp}\Biggl[-{1\over {2\beta_{\parallel}}}\sum_{l=1}^{N}
\sum_{\vec r} \vec n^2(\vec r,l)
-{1\over{2\beta_{\perp}}}\sum_{l=1}^{N-1}
\sum_{\vec r}n_z^2(\vec r,l)\Biggr],
\eqno (2)$$
where $n_{\mu}(r)$ is  an integer link-field
on the layered lattice structure  of
points $r=(\vec r, l)$, with   $\mu=x,y,z$ and
$\vec n=(n_x,n_y)$.
In the context of  the dual systems
of layered anyon superconductors
and spin-liquids, $\vec n(\vec r,l)$ represents the
confining statistical electric
field experienced between particle-hole excitations
about such groundstates.$^6$  Here one must make the
identification  $\beta_{\perp,\parallel}^{-1}
\rightarrow g_{\perp,\parallel}^2\hbar\omega_0/k_B T^{\prime}$,
where $g_{\perp,\parallel}$ denote the corresponding
gauge-field coupling constants, and where $T^{\prime}$
is the temperature in these dual systems.  The
Debye frequency scale, $\omega_0 = c_0 a^{-1}$,
is set by the characteristic velocity $c_0$,
which represents the
spin-wave velocity in the case of spin-liquids, and
the zero-sound speed in the case of
anyon superconductivity.$^{6}$  Note that the
case of weakly coupled planes, $J_{\perp}\ll J_{\parallel}$,
also corresponds to weakly electro-magnetically coupled
layers in the dual system, where
the coupling-constants satisfy
$g_{\perp}^{-2}\ll g_{\parallel}^{-2}$.

{\it Double-layer.}  Consider now the dual representation outlined
above for the special case of two coupled $XY$-model/compact QED
layers.  This problem may be solved exactly in
the limit of weak interlayer coupling, $J_{\perp}\ll J_{\parallel}$,
by first making
the decomposition $\vec n(\vec r,1)= \vec n^{\prime}
(\vec r, 1) -\vec n_{-}(\vec r)$ and $\vec n(\vec r,2)= \vec n^{\prime}
(\vec r, 2) +\vec n_{-}(\vec r)$, such that the intra-layer
fields $\vec n^{\prime}(\vec r, l)$ satisfy
$\vec\nabla\cdot  \vec n^{\prime}|_{r}=\sum_{r^{\prime}}
 p(r^{\prime})\delta_{r r^{\prime}}$,
while the inter-layer field $\vec n_{-}(\vec r)$ satisfies
$\vec\nabla\cdot\vec  n_{-}|_{\vec r}=n_z(\vec r)$,
with $\vec\nabla = (\Delta_x, \Delta_y)$.
We now take the customary potential
representation $\vec n_{-}=-\vec \nabla\Phi$
for the inter-layer field,
which yields
$\Phi(\vec r) =
\sum_{\vec r\,^{\prime}} G^{(2)}(\vec r - \vec r\,^{\prime})
n_z(\vec r\,^{\prime})$, where $G^{(2)}(\vec r)= (2\pi)^{-2}
\int_{\rm BZ}d^2k (e^{i\vec k\cdot \vec r}-1)
[4-2 {\rm cos} (k_x  a) -2 {\rm cos} (k_y a)]^{-1}$ is the Greens function
for the square lattice.  Note that fluxon-charge neutrality is
presumed;$^9$ i.e., $\sum_{\vec r} n_z(\vec r) = 0$.
The
requirement that the original
field $\vec n$ be integer valued
indicates that the intra-layer
field $\vec n^{\prime}$
is  not integer-valued, in general.
However, in the presently considered limit
of weak coupling, $J_{\perp}/J_{\parallel}\rightarrow 0$,
the concentration of inter-layer (fluxon) charge, $n_z$,
is exponentially small.  This  implies that
the inter-layer field, $\vec n_-$,
is  equally
small in magnitude, and that the intra-layer field,
$\vec n^{\prime}$, is
very close to being integer valued.  After
making a  suitable (lattice)
integration by parts of the energy functional in Eq. (2),
we thus obtain the factorization $Z=Z_{\rm DG}^{(1)}
Z_{\rm DG}^{(2)}Z_{\rm CG}$
for the partition function in the limit of
weak coupling between layers, where
the intra-layer factors represent 2D discrete gaussian  models
that  read
$$Z_{\rm DG}^{(l)}[p] = \sum_{\{\vec n^{\prime}(\vec r, l)\}}\Pi_{\vec r}
\delta\Bigl[\vec\nabla\cdot\vec n^{\prime}|_{\vec r, l}
-\sum_{\vec r\,^{\prime}} p(\vec r\,^{\prime}, l)
\delta_{\vec r\,^{\prime}, \vec r}\Bigr]
{\rm exp}\Biggl[-{1\over {2\beta_{\parallel}}}
\sum_{\vec r} \vec n^{\prime 2}(\vec r,l)\Biggr],\eqno (3)$$
and where the inter-layer factor
is given by the
2D Coulomb gas ensemble
$$\eqalignno{Z_{\rm CG}[p]=&\sum_{\{n_z(\vec r)\}} {\rm exp}
\Biggl\{-{1\over {2\beta_{\parallel}}} \sum_{\vec r}
2 [ n_z(\vec r)- p(\vec r, 1) + p(\vec r, 2)]\Phi(\vec r)
-{1\over{2\beta_{\perp}}}\sum_{\vec r} n_z^2(\vec r)\Biggr\}\cr
=&\sum_{\{n_z(\vec r)\}} {\rm exp}
\Biggl\{-{1\over {2\beta_{\parallel}}} \sum_{\vec r, \vec r\,^{\prime}}
2 [ n_z(\vec r)- p(\vec r, 1) + p(\vec r, 2)]G^{(2)}(\vec r-\vec r\,^{\prime})
n_z(\vec r\,^{\prime}) - {1\over{2\beta_{\perp}}}\sum_{\vec r}
n_z^2(\vec r)\Biggr\}.\cr
& &(4)\cr}$$
Considering first the case of absent probe charges, $p(r)=0$,
we immediately see that each layer (3)
undergoes a pure 2D $XY$/discrete-gaussian
model transition at $k_B T_c\lsim {\pi\over 2} J_{\parallel}$,
while the inter-layer links $n_z(\vec r)$ undergo an
inverted  2D Coulomb gas
binding transition at $k_B T_*= 4\pi J_{\parallel}$ in the limit
of weak inter-layer coupling, $J_{\perp}\ll J_{\parallel}$.  We therefore
find that Korshunov's approximate results for the corresponding
problem of infinitely many layers are {\it exact} in the present instance
of double layers.$^{12}$  Following this author,
the latter high-temperature
transition therefore corresponds to an inter-layer decoupling
transition mediated by the binding of vortex rings  lying in between
the double-layers.

Next consider the intra-layer phase autocorrelation
function,
$C_{11}(R)=\langle {\rm exp}[i\phi(0, 1)-i\phi(\vec R, 1)]\rangle$,
 for the $XY$-model, which corresponds
to fixing  two opposing unit charges
separated by a distance $R$ within the dual
compact QED representation; e.g., we have
$p(\vec r, 1) = \delta_{\vec r, 0} - \delta_{\vec r,\vec R}$
and $p(\vec r, 2) = 0$.
By the previous factorization for the partion
function, we may therefore write
$C_{11}(R)=C_{\parallel}(R) C_{\perp}(R)$,
where $C_{\parallel}(R)=Z_{\rm DG}^{(1)}[p]/Z_{\rm DG}^{(1)}[0]$
and $C_{\perp}(R)=Z_{\rm CG}[p]/Z_{\rm CG}[0]$.  The ``parallel''
 autocorrelation
is given by that of the pure square-lattice $XY$ model;$^9$
i.e., $C_{\parallel}(R)=
(r_0/R)^{\eta_{\parallel}} e^{-R/\xi_{\parallel}}$, where the
correlation length diverges as $\xi_{\parallel}/a
\sim {\rm exp}[A/(T/T_c -1)^{1/2}]$ just above $T_c$,$^8$
 and where the exponent is given by
$\eta_{\parallel}=(2\pi\beta_{\parallel})^{-1}$ at low-temperature
and  by $\eta_{\parallel}=1/4$ at $T_c$.
Here, ${\rm ln}\, r_0^{-1}=1.6169$.
To compute the ``perpendicular'' factor $C_{\perp}(R)$ to the
intra-layer correlation function, we first re-express the
right side of Eq. (4) as
$Z_{\rm CG}[p]/Z_{\rm CG}[0]={\rm exp}\{
[G_{\rm MF}^{(2)}(R)
-G^{(2)}(R)]/
2\beta_{\parallel}
\}$,
where the meanfield renormalized Coulomb interaction
is defined by$^{7}$
$$\eqalignno{G_{\rm MF}^{(2)}(R)
=2\beta_{\parallel} {\rm ln}
\Biggl[(Z_{\rm CG}[0])^{-1}
\sum_{\{n_z(\vec r)\}} {\rm exp}
\Biggl\{& -{1\over {2\beta_{\parallel}}} \sum_{\vec r, \vec r\,^{\prime}}
 2\Bigl[n_z(\vec r)-
{1\over 2}\bigl(\delta_{\vec r, 0}-\delta_{\vec r, \vec R}\bigr)\Bigr]
G^{(2)}(\vec r-\vec r\,^{\prime})\times\cr
&\times\Bigl[n_z(\vec r\,^{\prime})-
{1\over 2}\bigl
(\delta_{\vec r\,^{\prime}, 0}-\delta_{\vec r\,^{\prime}, \vec R}\bigr)\Bigr]
- {1\over{2\beta_{\perp}}}\sum_{\vec r}
n_z^2(\vec r)\Biggr\}\Biggr].\cr
& &(5)\cr}$$
Therefore, for $T<T_*$ the above meanfield renormalized Coulomb interaction
is short-ranged, yielding
$C_{\perp}(R)={\rm exp}[
-G^{(2)}(R)/
2\beta_{\parallel}]
\cong (R/r_0)^{1/4\pi\beta_{\parallel}}$, while it is
dielectrically renormalized
to $G_{\rm MF}^{(2)}(R)=\epsilon^{-1}(T) G^{(2)}(R)$ at
$T>T_*$, yielding
$C_{\perp}(R)
\cong  (R/r_0)^{(1-\epsilon^{-1})/4\pi\beta_{\parallel}}$.
Putting together all of these results, we obtain
$C_{11}(R)=(r_0/R)^{\eta}e^{-R/\xi_{\parallel}}$
for the inter-layer correlator,
where the temperature dependent exponent is given by
$$\eqalignno{
           &\qquad {1\over{4\pi\beta_{\parallel}}}; \quad T\ll T_c\cr
      \eta   =     &\qquad {1\over 8}; \quad T\sim T_c & (6)\cr
           &\qquad \eta_{\parallel}-{1-\epsilon^{-1}
\over{4\pi\beta_{\parallel}}};
\quad T> T_*.\cr}$$
Notice that
the above correlations are strengthened
by the second layer
with respect
to those of a single layer
below the decoupling transition $T_*$,
while they approach those
of an isolated single layer at high-temperature
[${\rm lim}_{T\rightarrow\infty}\epsilon(T) = 1$].

To compute the inter-layer phase auto-correlation function
$C_{12}(R)=\langle {\rm exp}[i\phi(0, 1)-i\phi(\vec R, 2)]\rangle$,
it becomes useful to consider the corresponding probe charges
as a superposition of the previous intra-layer ones
along with the minimum-distance  inter-layer
configuration $p^{\prime}(\vec r, 1)= \delta_{\vec r, \vec R}$
and $p^{\prime}(\vec r, 2) = -\delta_{\vec r, \vec R}$.
The latter configuration can be incorporated
into the  inter-layer field $\vec n_{-}(\vec r)$, resulting in
the modified constraint
$\vec\nabla\cdot\vec n_{-}|_{\vec r}=n_z(\vec r)
-\delta_{\vec r, \vec R}$.  After appropriately modifying
the manipulations
beginning with Eq. (2), we ultimately obtain
$C_{12}(R)=
C_{\parallel}(R)
{\rm exp}\{-[G^{(2)}(R)+G_{\rm MF}^{(2)}(R)]/2\beta_{\parallel}\}
=(r_0/R)^{\eta_{\parallel}-(1+\epsilon^{-1})/4\pi\beta_{\parallel}}
e^{-R/\xi_{\parallel}}$ for the
inter-layer correlation function.  Notice, therefore, that
this inter-later correlator  is {\it equal} to the intra-layer
correlator, $C_{11}(R)$, at temperatures
below the decoupling transition, $T < T_*$, where  $\epsilon^{-1}(T)=0$.
We see explicitly, therefore, that the double-layers
become locked at temperatures below $T_*$.

In terms of double-layer compact QED, which describes the gauge-field
dynamics in  both double-layer anyon superconductors
and spin-liquids,$^6$  the above discussion implies that there
exists a deconfinement transition at $k_B T_c^{\prime}\gsim
{2\over{\pi}}g_{\parallel}^2\hbar\omega_0$, below which
both intra-layer [$C_{11}(R)$] and inter-layer [$C_{12}(R)$]
particle-hole excitation in either system are confined
by a statistical electric flux tube$^{10}$ with string tension
$\gamma(T^{\prime}) = k_B T^{\prime}/\xi_{\parallel}(T^{\prime})$.
Note that this result is identical to that
of a single layer.  The present double-layer case, however,
also shows a
decoupling transition at $k_B T_*^{\prime} = (4\pi)^{-1} g_{\parallel}^2
\hbar\omega_0\sim k_B T_c^{\prime}/10$  between
the corresponding
in-plane statistical electric fields, $\vec n (\vec r, l)$.
In particular, the Fourier transform of
the inter-layer correlation function for
this field is given by
$$\langle\vec n(\vec k, 1)\cdot\vec n(-\vec k, 2)\rangle
 = - \langle\vec n_-(\vec k)\cdot\vec n_-(-\vec k)\rangle
 = - {k^2\over{\epsilon_0 (k^2+\xi_{\perp}^{-1})}},\eqno (7)$$
where $\xi_{\perp}(T^{\prime})$ denotes the Debye
screening length for the 2D Coulomb gas of fluxons.
Above, we have used the potential representation
$\vec n_- = - \vec\nabla\Phi$ for the ``difference''
field, along with the identity
$\langle\Phi(\vec k)\Phi(-\vec k)\rangle = \epsilon_0^{-1}
(k^2+\xi_{\perp}^{-1})^{-1}$ for
 the Coulomb gas correlation
function. Hence, the homogeneous inter-layer correlation
$\langle\vec n(0, 1)\cdot\vec n(0, 2)\rangle$ jumps
from $-\epsilon_0^{-1}(T_*^{\prime})=-{2\over\pi}$
to zero upon heating through decoupling transition-temperature
$T_*^{\prime}$.
This indicates that the deconfinement transition
at $T_c^{\prime}>T_*^{\prime}$ is 2D, since the layers
have already become decoupled by then.
Finally, we note that the specific-heat of  the  present double-layer
model for strong-coupling compact QED naturally decouples
into $C_v = 2 C_{\rm DG} + C_{\rm CG}$, where $C_{\rm DG}$ and $C_{\rm CG}$
denote the  contributions due to an     isolated
layer (3) described by the discrete-gaussian model$^{6}$
and those due
to the inter-layer Coulomb-gas ensemble     (4),
respectively.  Given that
$C_{\rm CG}(T^{\prime})$ shows a
smooth peak just above$^8$ $T_*^{\prime}$,
while $C_{\rm DG}(T^{\prime})$ exhibits one just below$^6$
$T_c^{\prime}\sim 10 T_*^{\prime}$, we expect that the
total specific-heat of the double-layer, $C_v(T^{\prime})$,
has two smooth peaks in the regime $T_*^{\prime} < T^{\prime}
< T_c^{\prime}$.

$N$-{\it layers.}  Let us now generalize the previous
discussion to the  case of $N$ weakly coupled $XY$
models (1).
To simplify matters, let us also consider the
partition function (2)
in the absence of probe charges, i.e., $p=0$.
The dual in-plane
field can then be expressed as
$\vec n(\vec r, l) = \vec n\,^{\prime}(\vec r, l)
- \vec n_-(\vec r, l) + \vec n_-(\vec r, l-1)$,
along with constraints
$\vec\nabla\cdot\vec n^{\prime} = 0$ and
$\vec\nabla\cdot\vec n_- = n_z$.  Again, taking
the customary potential form for the
``difference'' field, $\vec n_- = -\vec\nabla\Phi$,
and making suitable (lattice) integration by parts,
one obtains the factorization
$Z = Z_{\rm CG}\Pi_{l=1}^{N-1} Z_{\rm DG}^{(l)}$ for
the partion function in the limit
$J_{\perp}/J_{\parallel}\rightarrow 0$, where the Coulomb gas portion (4)
now has the generalized form
$$\eqalignno{
Z_{\rm CG} =  \sum_{\{ n_z(\vec r, l)\}} {\rm exp}
\Biggl\{ -{1\over{2\beta_{\parallel}}}\sum_{l=1}^{N}
\sum_{\vec r}[\vec n_-(\vec r, l-1) - \vec n_-(\vec r, l)]&^2
-{1\over{2\beta_{\perp}}}\sum_{l=1}^{N-1}
\sum_{\vec r} n_z^2(\vec r, l)\Biggr\} \cr
       =  \sum_{\{ n_z(\vec r, l)\}} {\rm exp}
\Biggl\{ -{1\over{2\beta_{\parallel}}}\sum_{l=1}^{N}
\sum_{\vec r, \vec r\,^{\prime}}[n_z(\vec r, l-1) - n_z(\vec r, l)] &
G^{(2)}(\vec r - \vec r \,^{\prime})
[n_z(\vec r\,^{\prime}, l-1) - n_z(\vec r\,^{\prime}, l)]\cr
&-{1\over{2\beta_{\perp}}}\sum_{l=1}^{N-1}
\sum_{\vec r} n_z^2(\vec r, l)\Biggr\},& (8) \cr}$$
with the fields at the boundary layers set to
$n_z(\vec r, 0) = 0 = n_z(\vec r, N)$.
The key difference between the present Coulomb gas
ensemble and the elementary one associated with
the double-layer case (4) is that consecutive vertical
fields, $n_z(\vec r, l-1)$ and $n_z(\vec r, l)$,
of equal value make no contribution to the energy
functional.  However, in the limit of weakly
coupled planes, Eq. (8) indicates that the concentration
of dipolar (fluxon) charge excitations between consecutive
planes is exponentially low.  This observation
suggests that each set of consecutive layers decouple
precisely like the isolated  double-layer at $T_*$,
in agreement with the results obtained by Korshunov
in his approximate treatment of the
corresponding problem of infinitely many layers.$^{12}$
At temperatures below $T_*$, we expect that
consecutive layers will be locked, with each layer
in turn
attaining quasi long-range order at
temperatures below $T_c\sim T_*/10$.  In this picture,
therefore, the superfluid BKT transition involves the
entire layered structure, making it 3D-like.$^{13}$

The above   picture may fail, however, in the
regime of moderately coupled to strongly coupled
planes, where the assumption that dipolar (fluxon) charge
excitations be  dilute is no longer valid.
This is certainly true for the
case of the isotropic problem with infinitely many layers
(3D $XY$ model), where the decoupling transition
necessarily must be
driven to infinite temperature due to the occurrence of a
{\it single} 3D phase transition in the system.
Hikami and Tsuneto have argued that vortex rings traversing
many layers have a minimum mean radius given by
$r \sim (J_{\parallel}/J_{\perp})^{1/2}$ in units of
the lattice constants.$^{14}$
Therefore, comparison of the latter to the total number
of layers, $N$, yields a minimum anisotropy condition
$J_{\parallel}/J_{\perp}\gsim N^2$ that insures
the existence of the BKT
superfluid transition
characteristic of weakly coupled double-layers.
This indicates, however, that the $XY$ model
with infinitely many layers
may  {\it not} be in the same universality class as that of
the weakly coupled double-layer, contrary to claims
in the literature.$^{12}$

In conclusion, we find that the superfluid transition exhibited
by a finite number (two or greater) of weakly coupled
square-lattice $XY$ model layers
in the dual Villain form is 3D-like$^{13}$ while
remaining in the same universality class as that of the 2D
$XY$ model.$^{12}$  This result suggests that
the latter transition is driven
not by the unbinding of isolated vortex-antivortex
pairs
in a single layer, which are in fact (linearly) confined,$^{14,15}$
but by the unbinding of {\it lines} of vortex-antivortex pairs
passing perpendicularly through all of the layers.
We also find an inter-layer decoupling transition in the
normal state at much
higher temperature.$^{12}$
Last, in terms of the dual problems
of lattice anyon superconductors and spin-liquids,$^6$
these results indicate that the essential nature
of the respective superconductor/normal
and spin-liquid/paramagnetic
transitions found in 2D is preserved in the present
case  of weakly coupled layered structures.

Discussions with M. Gabay, M. Maley, J. Engelbrecht
 and S. Trugman are gratefully
acknowledged.    This work
was performed under the auspices of the U.S. Department of Energy,
and  was supported in part by
the Associated Western Universities and  National
Science Foundation grant DMR-9322427.

\vfill\eject
\centerline{\bf References}
\vskip 16 pt

\item {1.} {\it The Physical Properties
of High-Temperature Superconductors},
vol. 2, edited
by D.M. Ginsberg (World Scientific, Singapore, 1990).

\item {2.}  P.W. Anderson, Science
 {\bf 235}, 1196 (1987).

\item {3.} R.B. Laughlin, Phys. Rev. Lett. {\bf 60}, 2677 (1988);
R.B. Laughlin, Science {\bf 242}, 525 (1988).

\item {4.}  J.P. Rodriguez and B. Dou\c cot,
Phys. Rev. B{\bf 45}, 971 (1992).

\item {5.} N. Read and S. Sachdev, Phys. Rev. Lett. {\bf 62}, 1694
(1989); J.P. Rodriguez, Phys. Rev. B {\bf 41},
7326 (1990); A. Sokol and D. Pines, Phys. Rev. Lett.
{\bf 71}, 2813 (1993).

\item {6.} J.P. Rodriguez, Los Alamos Report \# LA-UR-94-3002 (1994).

\item {7.} S.T. Chui and J.D. Weeks, Phys. Rev.
B {\bf 14}, 4978 (1976).

\item {8.} P. Minnhagen, Rev. Mod. Phys. {\bf 59}, 1001 (1987).

\item {9.}  J.V. Jos\' e,
L.P. Kadanoff, S. Kirkpatrick and
 D.R. Nelson, Phys. Rev. B {\bf 16},
 1217 (1977);  C. Itzykson and J.
 Drouffe, {\it Statistical field theory},
vol. 1, chap. 4 (Cambridge Univ.
Press, Cambridge, 1991).

\item {10.} A. Polyakov, Phys. Lett.  {\bf 72B}, 477 (1978);
L. Susskind, Phys. Rev. D {\bf 20}, 2610 (1979).

\item {11.} X.G. Wen, F. Wilczek, and A. Zee, Phys. Rev. B {\bf 39},
11413 (1989).

\item {12.} S.E. Korshunov, Europhys. Lett. {\bf 11}, 757 (1990).

\item {13.} S.T. Chui and M.R. Giri, Phys. Lett. A {\bf 128},
49 (1988).

\item {14.} S. Hikami and T. Tsuneto, Prog. Theor. Phys. {\bf 63},
387 (1980).

\item {15.} L.I. Glazmann and A.E. Koshelev, Zh. Eksp. Teor. Fiz.
{\bf 97}, 1371 (1990) [Sov. Phys. JETP {\bf 70}, 774 (1990)];
V. Cataudella and P. Minnhagen, Physica C {\bf 166}, 442
(1990); P. Minnhagen and P. Olsson, Phys. Rev. B {\bf 44},
4503 (1991).


\end

\item {9.} A. Fetter, C. Hanna,
R.B. Laughlin, Phys. Rev. B {\bf 39},
9679 (1989).

\item {10.} Y. Chen, F. Wilczek,
E. Witten, and B.I. Halperin, Int. J.
Mod. Phys. B {\bf 3}, 1001 (1989).

\item {11.} P.B. Wiegmann, Phys. Rev. Lett. {\bf 65}, 2070 (1990);
J.P. Rodriguez. and B. Dou\c cot,
Phys. Rev. B{\bf 42}, 8724 (1990); (E) {\bf 43}, 6209 (1991).

\item {13.} L.B. Ioffe and A.I. Larkin,
 Phys. Rev. B {\bf 39}, 8988 (1989).

\item {16.} I. Ichinose and T. Matsui, Nucl. Phys.
{\bf B394}, 281 (1993).

\item {17.} N. Nagaosa, Phys. Rev. Lett. {\bf 71},
4210 (1993).

\item {18.} J.W. Loram et al., Phys. Rev. Lett. {\bf 71},
1740 (1993);
H. Wuhl et al., Physica C {\bf 185 - 189}, 755 (1991);
A. Junod in ref. 1.

\item {19.}  A.A. Abrikosov, L.P. Gorkov,
and I.E. Dzyaloshinski, {\it Methods of
Quantum Field Theory in Statistical Physics},
(Dover, New York, 1975).

\item {20.} A.M. Polyakov, Nucl. Phys. B{\bf 120}, 429 (1977).

\item {21.} A.M. Polyakov, {\it Gauge Fields and Strings}
(Harwood, New York, 1987).

\item {22.} T. Banks, R. Myerson and J. Kogut, Nucl. Phys. {\bf B129},
493 (1977).

\item {25.} R.H. Swendsen, Phys. Rev. B {\bf 15}, 5421 (1977).

\item {26.} H. van Beijeren and I. Nolden, in
{\it Structure and Dynamics of Surfaces II},
edited by W. Schommers
and P. von Blanckenhagen (Springer, Heidelberg, 1987).

\item {27.} J.E. Hetrick, Y. Hosotani, and B.H. Lee,
Ann. Phys. (NY) {\bf 209}, 151 (1991).

\vfill\eject
\centerline{\bf Figure Caption}
\vskip 20pt
\item {Fig. 1.}

\end